\documentstyle[epsf,epsfig]{JFM}

\title[Amplitude equations and pattern selection ...]
{Amplitude equations and pattern selection in Faraday waves}
\author[P. Chen and J. Vi\~nals]{PEILONG CHEN$^{1}$ AND JORGE 
VI\~NALS$^{1,2}$}

\affiliation{$^1$Supercomputer Computations Research Institute,
             Florida State University, Tallahassee, Florida 32306-4130, USA\\
             $^{2}$ Department of Chemical Engineering,
             FAMU-FSU College of Engineering, Tallahassee, Florida
		31310-6046, USA}
\pubyear{1998}
\volume{000}
\pagerange{000--000}
\date{\today}
\begin{document}
\maketitle

\begin{abstract}
A nonlinear theory of pattern selection in
parametric surface waves (Faraday waves) is presented that is
not restricted to small viscous dissipation. By using a multiple scale
asymptotic expansion near threshold, a standing wave amplitude
equation is derived from the governing equations.
The amplitude equation is of gradient
form, and the coefficients of the associated Lyapunov function 
are computed for regular patterns of various symmetries as a function
of a viscous damping parameter $\gamma$.
For $\gamma \sim 1$, the selected wave pattern comprises a single standing 
wave (stripe pattern). For $\gamma \ll 1$,
patterns of square symmetry are obtained in the capillary
regime (large frequencies).  At lower frequencies (the mixed
gravity-capillary regime), a sequence of six-fold (hexagonal),
eight-fold, $\ldots$ patterns are predicted. For even lower frequencies
(gravity waves) a stripe pattern is again selected. Our predictions of the
stability regions of the various patterns are in quantitative
agreement with recent experiments conducted in large aspect ratio
systems.
\end{abstract}

\section{Introduction}

This paper extends an earlier calculation by \cite{re:zhang97} of the amplitude
equation governing Faraday waves in the weakly nonlinear regime.
In order to make the
problem analytically tractable, they neglected without rigorous 
justification viscous terms in the boundary conditions at the free fluid
surface
that had a nonlinear dependence on either the surface displacement away from
planarity, or on the surface velocity. 
Even though the resulting amplitude
equation led to the prediction of stationary patterns that are
generally in agreement with experiments conducted in the regime of weak viscous
dissipation (\cite{re:kudrolli96a,re:binks97}), the unsystematic nature
of the truncation makes it difficult to asses the
range of validity of the theory. In particular, the so-called stripe pattern
(a pattern comprised of a single standing wave) which is generically observed
when viscous dissipation is not small, could not be obtained in their
analysis for any range of parameters. We extend below this earlier work,
and present a systematic weakly nonlinear theory of Faraday waves. Our results 
on pattern selection agree with those of
\cite{re:zhang96}  and (1997) in the limit of small viscous dissipation, and 
with recent experimental work otherwise.

Parametrically driven surface waves (also known as Faraday waves)
can be excited on the free surface of a fluid layer 
that is periodically 
vibrated in the direction normal to the surface at rest 
when the amplitude of the driving acceleration is large
enough to overcome the dissipative effect of fluid viscosity
(\cite{re:faraday31,re:miles90}). Of special concern to us is the issue of
pattern selection in a layer of lateral
dimension much larger than the excited wavelength (see, e.g., 
\cite{re:cross93} for a recent review on pattern formation). 
In the case of Faraday waves, it is now known
that different wave patterns can be excited depending on the fluid properties 
and the driving amplitude or frequency. At high viscous dissipation
(a fluid of large viscosity and/or a low driving frequency),
the observed wave pattern above threshold consists of parallel
stripes (\cite{re:edwards94} and \cite{re:daudet95}). For
lower dissipation, patterns of square symmetry (combinations of two
perpendicular plane waves) are observed in the capillary regime (large
frequencies) 
(\cite{re:lang62,re:ezerskii86,re:tufillaro89,%
re:ciliberto91}
\hfil\break
\cite{re:christiansen92,re:muller93,re:edwards94}). 
At lower frequencies (the mixed
gravity-capillary regime), higher symmetry 
patterns have been observed by \cite{re:kudrolli96a} (hexagonal)
and \cite{re:binks97} (hexagonal, eight- and ten-fold). The
aim of this paper is to present a weakly nonlinear analysis of Faraday waves
that predicts stationary wave patterns with these symmetries.

The derivation of an amplitude equation is a classical method to
describe excited states beyond linear instability. Just above
threshold, the evolution of the system is assumed to be described in terms 
of the complex amplitude $A$ of the most unstable mode according to linear 
theory. The equation of motion for $A$ is often of the form
\begin{equation}
  {dA\over dt} = \alpha A - g A^3,
  \label{eq:amp_intro}
\end{equation}
where $\alpha$ is the linear growth rate, and $g > 0$ is real. The low order 
nonlinear term provides saturation. There exist cases,
however, in which spatial isotropy permits waves to be excited in any
direction, and the nonlinear interaction term in the equation above 
contains terms of the form
$g_{ij}|A_j|^2A_i$, with $A_i$ and $A_j$ the slowly varying
amplitudes of two degenerate unstable modes. If the coupling coefficients 
$g_{ij}$ are known, the resulting wave pattern can be predicted from
Eq. (\ref{eq:amp_intro}), as has been illustrated by \cite{re:muller94} 
for Faraday waves.

The derivation of amplitude equations for surface waves
is greatly simplified in the case of an
ideal (inviscid) fluid. Since the bulk flow is irrotational,
there exists a hamiltonian formulation in which 
the canonically conjugate
variables are the surface displacement and the velocity potential at the free
surface (\cite{re:zakharov68,re:miles77}). As a consequence,
early analyses of Faraday waves were based on the
hamiltonian description of the inviscid limit, and treated
viscous or dissipative effects as a perturbation
(\cite{re:miles84,re:milner91,re:miles93}). The derivation usually starts 
from the set of ideal fluid equations (\cite{re:benjamin54}), written 
in terms of the surface velocity potential $\phi$. The linear or zero-th order
solution $\phi_{0}$ is a sum over waves of frequency $\omega$ and wavevector
$\{{\bf k}_j\}$, with $\omega$ and $k= \| {\bf k}_{j} \|$ related by the ideal 
fluid dispersion 
relation (Eq. (\ref{eq:ideal_dispersion}) below):
$$
  \phi_0 = {-i\omega\over k}e^{kz}\sum_j A_j(T)e^{i({\bf k}_j\cdot
  {\bf x}-\omega t)} + \hbox{c.c.},
$$
where $T = \epsilon t$ is a slow time scale, with $\epsilon \ll 1$ the
dimensionless distance away from threshold. An expansion of the ideal
fluid equations to third order in $\epsilon$ yields the equation for 
$A_{j}(T)$ (\cite{re:milner91})
\begin{equation}
\label{eq:milner}
  {dA_j\over dT} = -{ikf\over 4\omega}A^*_{-j} 
         +i\sum_k\Pi^{(1)}_{jk}|A_k|^2A_j
         +i\sum_k\Pi^{(2)}_{jk}A_kA_{-k}A^*_{-j},
  \label{eq:ideal_amp_eq}
\end{equation}
with $f$ the amplitude of the driving acceleration,
and $\Pi$ {\em real} functions of the 
angle between the $j$-th and $k$-th wavevectors. Since the coefficients of the
cubic terms are imaginary, these terms do not contribute to wave saturation.
This can be seen, for example, by computing $d |A_{j}|^{2}/dT$ and observing
that all cubic terms cancel. This is also a manifestation of a general
symmetry principle in hamiltonian (or reversible) systems
that prohibits {\em real} coefficients of cubic nonlinear terms in
standing wave amplitude equations (\cite{re:cross93,re:coullet94}).

In the limit of low viscous dissipation, Hamilton's equations have been
supplemented with a dissipation function 
(\cite{re:miles84,re:milner91,re:miles93}),
which is computed under the assumption that the dominant contribution to
viscous dissipation arises from the irrotational velocity field in the bulk, 
and not from friction at the
container walls or dissipation near the free surface (where vorticity is
produced). Under this assumption, the rate of energy loss is given by 
(\cite{re:landau59}),
$$
  \dot E = -2\eta\int dV\left({\partial^2\phi\over\partial x_i\partial
   x_j}\right)^2,
$$
where $\eta$ is the shear viscosity, and the integral extends over the bulk 
fluid. The velocity potential $\phi$ is now expanded in powers of $\epsilon$,
and viscous contributions computed order by order in $\epsilon$. This procedure
leads to imaginary components in the coefficients of the cubic terms of
Eq. (\ref{eq:milner}),
and therefore to wave saturation. The precise functional form of the
coefficients obtained by this method is still somewhat controversial
(\cite{re:miles93,re:hansen97}).

We next address the effect of the rotational component of the 
flow. The dimensionless group involving the ratio of viscous to inertial
effects is the damping parameter $\gamma = 2 \nu
k_{0}^{2}/\omega_{0}$, where $k_{0}$ is the critical wavenumber 
in the inviscid limit, and
$\omega_{0}$ its angular frequency ($\gamma$ is inversely
proportional to the Reynolds number of the flow). \cite{re:lundgren88}
considered an expansion of the governing equations and boundary conditions
in powers of $\gamma$. They showed that in the weak dissipation limit,
the dominant terms in the boundary conditions are ${\cal O}(\gamma)$, with
a first correction at ${\cal O}(\gamma^{3/2})$. At {\em linear} order in 
the surface displacement or surface velocity, terms of ${\cal O}(\gamma)$
are purely irrotational, while the rotational flow component contributes
at ${\cal O}(\gamma^{3/2})$. In fact, linear stability analysis of Faraday 
waves by
\cite{re:muller97} (see also Sec. \ref{sec:equations}) shows that the
dimensionless value of the driving amplitude at threshold equals $\gamma$, with
a first correction term that is proportional to $\gamma^{3/2}$.
The dominant contribution arises solely from the irrotational flow component,
with contributions from the rotational component coming at
${\cal O}(\gamma^{3/2})$. However, \cite{re:zhang97} argued that 
the lowest order contribution from both irrotational and
rotational components is ${\cal O}(\gamma)$ 
at the nonlinear level in the surface variables.
Hence rotational flow cannot be neglected in a
nonlinear theory, even in the limit of small dissipation. For example,
the kinematic boundary condition at the free surface does
include one such term proportional to $\gamma$ that arises from the
component normal to the surface
of the rotational part of the velocity field. 
This term was retained both in the analysis
of \cite{re:zhang97} and in our analysis below,
but not in previous approaches based on a dissipation function.

Another qualitative feature of importance to pattern selection in Faraday 
waves is triad resonant interactions. Since the standing wave amplitude
equation must be invariant under $A_{j} \rightarrow -A_{j}$,
\footnote{The governing equations are invariant under time translation by a 
period of the driving force $t \rightarrow T + 2 \pi / \omega$. Subharmonic
response implies that $\zeta (x,y,t+2\pi / \omega) = - 
\zeta(x,y,t)$, with $z=\zeta(x,y,t)$ the position of the surface. Because of 
this invariance, the amplitude equation for $A_{j}$ must also be
invariant under a sign change in $A_{j}$}
triad resonance
cannot contribute directly to quadratic order in $A_{j}$, but it does
contribute significantly to the coefficients of the cubic order 
term via the coupling of the zero-th order unstable wave and 
first order stable waves.
This resonance was already encountered by \cite{re:milner91} as a
divergence of the cubic coefficient in the standing wave amplitude
equation at a particular angle. Later,
\cite{re:edwards94} suggested that triad resonance would be important at low
viscous dissipation, range in which linearly stable modes are only weakly
damped. \cite{re:zhang97} calculated such a contribution explicitly, and
showed that it is important in determining the symmetry of the selected 
pattern in the region of
small $\gamma$. In particular, they predicted a sequence of quasi-periodic
patterns in the region in which the resonant angle approaches zero.

We extend in this paper the analysis of \cite{re:zhang97} that was based
on a quasi-potential approximation to the governing equations.
By separating the rotational flow within a small vortical layer near the 
free surface from the
potential flow in the bulk, they derived a standing wave amplitude
equation valid in the limit of small viscous dissipation.
The calculation, however, relied on an uncontrolled approximation concerning
nonlinear viscous terms and, as a consequence, its region of validity
is difficult to asses. We describe below
a systematic expansion of the Navier-Stokes equation and boundary
conditions that overcomes this difficulty and that
leads to an amplitude equation not restricted to small
viscous dissipation. We start by deriving an exact, although implicit,
relation for the threshold of instability, which is then used in the
nonlinear analysis. This result extends earlier numerical work by
\cite{re:kumar94}, and agrees with a recent low viscosity approximation
to the governing equations by \cite{re:muller97}. We then use a multiple
scale expansion to derive a standing wave equation which is of gradient
form. Minimization of the associated Lyapunov function leads to the
prediction of stationary patterns of different symmetries
as a function of the fluid parameters and frequency of the driving
acceleration. Our predictions are in good agreement with experiments
conducted in large aspect ratio cells.
 
\section{Governing equations and linear stability}
\label{sec:equations}

We consider a semi-infinite fluid layer, unbounded in the $x-y$ direction,
extending to $z = -\infty$, and with a planar free surface at $z=0$ 
when at rest.
The fluid is assumed incompressible and Newtonian.
Under periodic vibration of the layer in the direction normal to the surface at
rest, the equation governing fluid motion (in the co-moving reference frame)
is
\begin{equation}
\label{eq:ns}
  \partial_t{\bf u} + ({\bf u}\cdot{\bf\nabla}){\bf u}
 = -\frac{1}{\rho}{\bf\nabla}p + \nu\nabla^2{\bf u} + g_{z}(t){\hat{\bf e}}_z,
\end{equation}
with ${\bf u}$ the velocity field, $p$ the pressure, $\rho$ and $\nu$ 
the density and kinematic viscosity of the fluid respectively, and
$ g_{z}(t) = - g - f\cos\omega t$
the effective gravity. \footnote{We use a driving acceleration
proportional to $\cos \omega t$ instead of $\sin \omega t$ as in
\cite{re:zhang97} to avoid, as discussed in that reference, the
complication related to the parity under time reversal of the driving
acceleration.}
The base state is a quiescent fluid with a pressure
distribution $p = \rho g_{z}(t)z$. We absorb the body force 
in the pressure, so that in what follows $p$ is the deviation from 
$\rho g_{z}(t)z$. By applying
$-\left( {\bf\nabla}\times{\bf\nabla}\times \right)$ to Eq. (\ref{eq:ns}), 
one can eliminate the pressure term, and
also obtain a system of equations for the velocity components of ${\bf u} = 
(u,v,w)$, in which the linear terms are uncoupled,
\begin{equation}
   \partial_t\nabla^2{\bf u} - \nu\nabla^2\nabla^2{\bf u}
  ={\bf\nabla}\times{\bf\nabla}\times({\bf u}\cdot{\bf\nabla}){\bf u}.
   \label{eq:momentum_equation}
\end{equation}
Continuity, ${\bf\nabla}\cdot{\bf u}=0$, has also been used to derive Eq.
(\ref{eq:momentum_equation}).

\begin{figure}
\epsfig{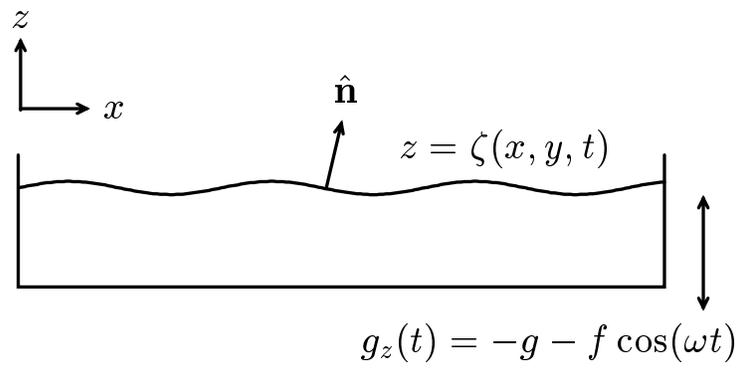}
  \vspace{-6cm}
  \caption{Schematic setup of a Faraday wave configuration.}
  \label{fig:setup}
\end{figure}

Besides the null conditions at $z=-\infty$, there are four boundary
conditions at the moving free surface (\cite{re:lamb45}). Let $z=\zeta(x,y)$
be the position of the surface
(Fig. \ref{fig:setup}), then the outward pointing unit normal ${\bf\hat n}$ is
\begin{equation}
  {\bf\hat n}={(-\partial_x\zeta,-\partial_y\zeta,1)\over
       [1 + (\partial_x\zeta)^2 + (\partial_y\zeta)^2]^{1/2}}.
  \label{eq:normal_vector}
\end{equation}
Two linearly independent tangential unit vectors ${\bf\hat t}_1$ and ${\bf\hat
t}_2$ are
\begin{equation}
 {\bf\hat t}_1=\frac{(1,0,\partial_x\zeta)}
                {[1 + (\partial_x\zeta)^2]^{1/2}}, \qquad
 {\bf\hat t}_2=\frac{(0,1,\partial_y\zeta)}
                {[1 + (\partial_y\zeta)^2]^{1/2}}.
 \label{eq:tangential_vector}
\end{equation}
Note that these two vectors are not mutually orthogonal. The
choice is made so that the expressions for ${\bf\hat t}_1$ 
and ${\bf\hat t}_2$ are symmetric in the Cartesian variables $x$ and $y$.

The kinematic boundary condition is,
$$
  \partial_t\zeta + \big({\bf u}(z=\zeta)\cdot{\bf\nabla}_H\big)\zeta =
  w(z=\zeta),
$$
with $\nabla_H =  {\bf\hat e}_x\partial_x + {\bf\hat
e}_y\partial_y$. Since the governing equations will be expanded and solved
order by order, we quote here its Taylor expansion around $z=0$,
\begin{equation}
  \partial_t\zeta + [u+\partial_zu\zeta]_{z=0}\partial_x\zeta
 +[v+\partial_zv\zeta]_{z=0}\partial_y\zeta
 =[w+\partial_zw\zeta+{\textstyle{1\over 2}}\partial_{zz}w\zeta^2]_{z=0}.
 \label{eq:kinematic}
\end{equation}
Only terms up to third order in the velocity or surface displacement
will be required. 
\footnote{In order to avoid excessive use of parentheses,
we follow the convention in the remainder of the paper that
the operator $\partial$ acts only on the function immediately following it.}

Neglecting the effect of the air phase above the fluid, the tangential
stress at the free surface is zero,
$$
  {\bf\hat t}_m\cdot{\bf T}\cdot{\bf\hat n}|_{z=\zeta} = 0, \qquad m=1,2,
$$
with ${\bf T}$ the stress tensor of components,
$ T_{ij} = [-p-\rho g_{z}(t)z]\delta_{ij}+\rho\nu(\partial_ju_i+\partial_iu_j). $
The normal stress at the fluid surface is balanced by capillarity,
$$ 
  {\bf\hat n}\cdot{\bf T}\cdot{\bf\hat n}|_{z=\zeta} = 2H\sigma,
$$
where $\sigma$ is the surface tension and $2H$ is the mean curvature of the 
surface,
$$
  2H=\left\{\partial_{xx}\zeta\left[1+(\partial_y\zeta)^2\right]
           +\partial_{yy}\zeta\left[1+(\partial_x\zeta)^2\right]
           -2\partial_x\zeta\partial_y\zeta\partial_{xy}\zeta\right\}\Big/
     \left[ 1+ (\partial_x\zeta)^2 + 
            (\partial_y\zeta)^2 \right]^{3/2}.
$$

The linear stability of the fluid layer under vibration was first addressed 
in the inviscid limit by \cite{re:benjamin54}, and later by \cite{re:landau76} 
in the limit of low viscosity. More recently,
\cite{re:kumar94} numerically computed the neutral stability curve for
a fluid of arbitrary viscosity, and \cite{re:muller97}
have given an analytic low viscosity expansion.
We first review briefly the formulation of \cite{re:kumar94}, and then
show that an exact (albeit implicit) analytical
expression for the threshold can be derived, thus avoiding the numerical
calculation.

The dominant response of the parametrically driven system is subharmonic
at a frequency $\omega /2$ (\cite{re:benjamin54}). Although the methodology 
discussed below can also be used to analyze a possible harmonic response, 
we restrict our analysis to the subharmonic case. To address the linear
stability of the fluid surface, we consider the following solutions for the
vertical velocity field and surface displacement,
\begin{eqnarray}
  w_0 &=& \cos(kx) \sum_{j=1,3,5,\cdots} e^{ji\omega t/2} w_{0}^j(z)A_j 
                          +\hbox{c.c.} \label{eq:linear_solution} \\
  \zeta_0 &=& \cos(kx) \sum_{j=1,3,5,\cdots} e^{ji\omega t/2} A_j
                          +\hbox{c.c.} \nonumber
\end{eqnarray}
where the $A_j$ are complex amplitudes, and we retain all the harmonics of
the fundamental mode $e^{i\omega t/2}$. Truncation of
the sums to include the fundamental mode $e^{i\omega t/2}$ alone is only 
appropriate for small viscous damping. From
Eq. (\ref{eq:momentum_equation}), the linearized equation of motion
for $w_0$ is
$$
   \left( \partial_t\nabla^2 - \nu\nabla^2\nabla^2 \right) w_0= 0.
$$
Substituting $w_0$ and $\zeta_0$ from (\ref{eq:linear_solution}),
one finds
$$
  \left[{\textstyle{1\over2}}ji\omega(-k^2+\partial_{zz})
      - \nu\left(-k^2+\partial_{zz}\right)^2\right]w_0^j(z) = 0.
$$
The solution of this equation is a linear combination of $e^{\pm kz}$ 
and $e^{\pm q_jz}$, with $q_j^2 =  k^2+ji\omega/2\nu$.
The linearized kinematic and tangential stress boundary conditions are
\begin{equation}
  \begin{array}{rcl}
     \partial_t\zeta_0 - w_0 &=& 0, \\
     (\nabla_H^2 - \partial_{zz})w_0 &=& 0.
  \end{array}
  \label{eq:zero_order_kin_tan}
\end{equation}
By using the boundary conditions (\ref{eq:zero_order_kin_tan}) and the
null conditions at $z=-\infty$, $w_0^j(z)$ is given by
$$
  w_0^j(z) = \left( \frac{1}{2} ji \omega + 2 \nu k^{2} \right)
e^{kz} - 2\nu k^2e^{q_j z}.
$$
The first term on the right hand side is the irrotational component of
the flow, in which we have explicitly separated the inviscid and viscous
contributions. The second term in the right hand side is the rotational
component (this is the component that has been neglected in earlier work by
\cite{re:milner91} and \cite{re:miles93}).
The linearized normal stress boundary condition is, after having eliminated
the pressure by using the equation of motion,
\begin{equation}
\left[ 2 \nu \nabla_H^2 - \left( \partial_t - \nu \nabla^{2} \right)
\right] \partial_zw_0
 + \left( g -\frac{\sigma}{\rho}\nabla_H^2 + f \cos\omega t
\right) \nabla_H^2\zeta_0=0.
\label{eq:w0linear}
\end{equation}
By substituting the assumed solutions given by
Eqs. (\ref{eq:linear_solution}) into this equation, we
note that the term $2 \nu \nabla_H^2 \partial_zw_0$ when acting on
the irrotational flow component $e^{kz}$ yields a contribution at low
viscosity that scales as $\nu$, whereas the rotational contribution
(from $e^{q_{j}z}$) scales as $\nu^{3/2}$. The remaining term 
$\left( \partial_t - \nu \nabla^{2} \right) \partial_zw_0$ is simply
equal to $- \omega^{2}$. Hence it is justified to 
neglect the rotational flow component in the linear stability analysis at low 
damping. As we show below, and in agreement with the work by
\cite{re:muller97}, rotational flow contributes terms of order $\nu^{3/2}$ and
higher to the value of the driving acceleration at onset. In the analysis that
follows, however, we will retain the full linear solution.

By equating the coefficients of each harmonic $e^{ji\omega t/2}$ resulting from
Eq. (\ref{eq:w0linear}), \cite{re:kumar94} found
\begin{equation}
  \begin{array}{rcl}
    H_1A_1 - fA_1^* - fA_3 &=& 0, \\
    H_3A_3 - fA_1 - fA_5 &=& 0, \\
    H_5A_5 - fA_3 - fA_7 &=& 0, \ldots \\
  \end{array}  \label{eq:eigen_zero}
\end{equation}
with
$$
  H_j = 2\left\{
      \nu^2\left[ 4q_jk^4-k(q_j^2+k^2)^2 \right] - gk^2 
        -{\sigma\over\rho}k^4\right\}\Big/k^2.
$$
This is a system of equations in the unknowns $A_{j}$, function of
wavenumber $k$ and driving amplitude $f$.
By truncating the system (\ref{eq:eigen_zero}) at some
particular $A_{n}$, it can be solved
numerically as an eigenvalue problem, $f$ being the eigenvalue.
This is indeed what was done by \cite{re:kumar94}. However we observe that after
truncation at $A_n$,
\begin{equation}
  A_n = {fA_{n-2}\over H_n}, \qquad
  A_{n-2} = {fA_{n-4} \over H_{n-2}-{f^2\over H_n}}, \qquad\ldots
  \label{eq:relation_An}
\end{equation}
so that the infinite set of equations can be re-written as
\begin{equation}
  \left( H_1 - { f^2 \over H_3 - {f^2 \over H_5- \cdots}}
  \right) A_1 - f A_1^* \equiv \bar H_1(k,f)A_1 - f A_1^* = 0.
  \label{eq:relation_AA}
\end{equation}
For a given wavenumber $k$, its threshold of instability $f_k$ is given
implicitly by
\begin{equation}
  f_k = |\bar H_1(k,f_k)| = \left| H_1 - {f_k^2\over H_3 - \cdots}\right|.
	\label{eq:threshold}
\end{equation}
The complex amplitude $A_j$ can then be recursively obtained from
Eq. (\ref{eq:relation_An}) and (\ref{eq:relation_AA}) up to a real
factor. The critical wavenumber for instability $k_{\hbox{\scriptsize
onset}}$ corresponds to the lowest value of $f_{k}$, $f_{0}$.

\begin{figure}
%  \epsffile{/home/scri18/users/peilong/papers/faraday_pattern_reg/fig2.ps}
\epsffile{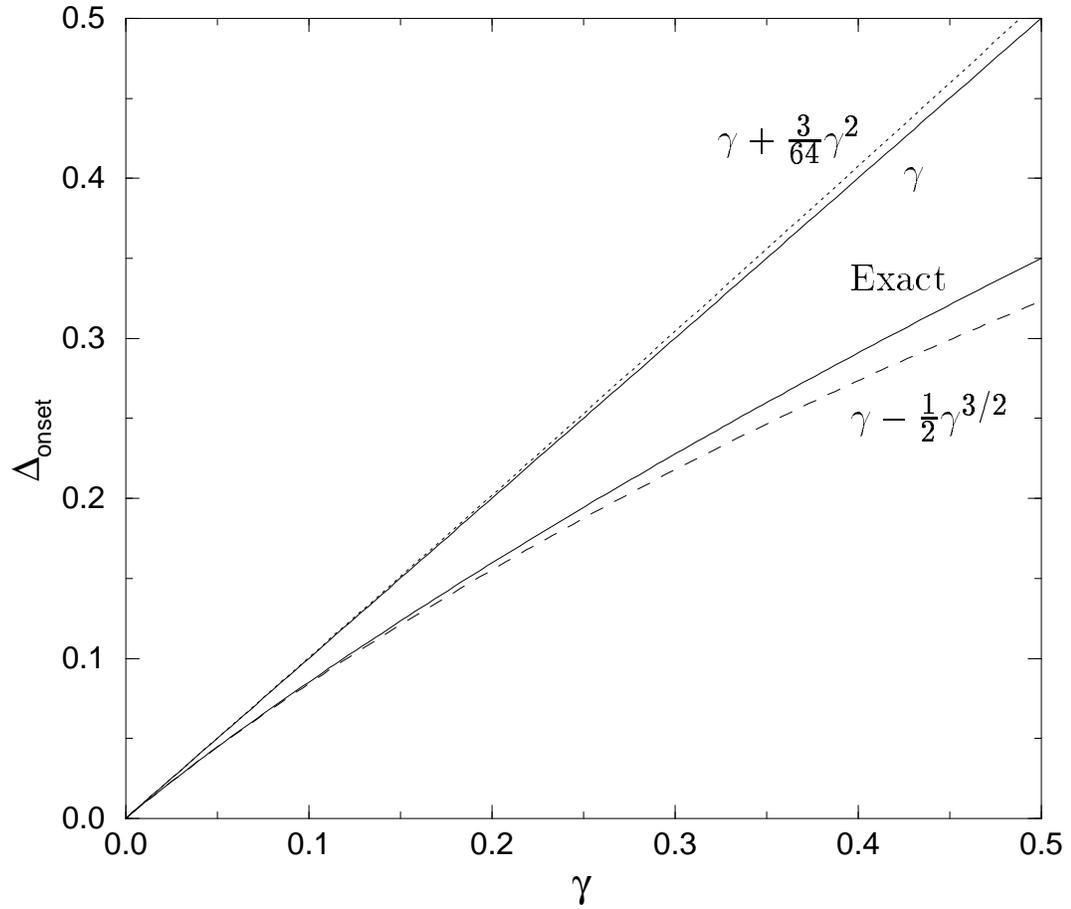}
  \caption{Dimensionless threshold for linear instability $\Delta_{\rm onset}$
  as a function of the dimensionless damping parameter $\gamma$. The lower 
  solid
  line is the exact result; the upper solid line is the lowest order
  approximation in the damping parameter. Also shown are the first order
  correction in the viscous damping parameter (dashed line), and
  the first correction for
  the instability threshold for a damped Mathieu equation.}
  \label{fig:threshold}
\end{figure}

It is interesting to consider the limiting behavior of 
Eq. (\ref{eq:threshold}) at low viscosity. First recall that for a 
semi-infinite inviscid fluid, the
dispersion relation for surface waves is (\cite{re:landau59})
\begin{equation}
  \omega_0^2 = gk_0 + \sigma k_0^3/\rho,
  \label{eq:ideal_dispersion}
\end{equation}
with $\omega_0 = \omega/2$, and $k_0$ the
wavenumber. In a fluid of low viscosity we expect
$k_{\hbox{\scriptsize onset}}$ to be near $k_0$.  It is then
convenient to define dimensionless variables by using $1/\omega_0$ as the
time scale, and $1/k_0$ as the length scale. We also define a reduced
wave number ${\bar k} = k/k_0$, a viscous damping coefficient $\gamma
= 2\nu k_0^2/\omega_0$, the gravity wave $G = gk_0/\omega_0^2$ and
capillary wave $\Sigma = \sigma k_0^3/\rho\omega_0^2$ contributions to
the dispersion relation, and the dimensionless amplitude of the driving
acceleration $\Delta = f k_0/4\omega_0^2.$ Note that
$G+\Sigma=1$ from Eq. (\ref{eq:ideal_dispersion}); $G=1$ corresponds
to a pure gravity wave while $G=0$ to a pure capillary wave. For
$\gamma\ll 1$, $k_{\hbox{\scriptsize onset}}$ and
$\Delta_{\hbox{\scriptsize onset}}$ in Eq. (\ref{eq:threshold}) can
be expanded as a power series of the damping coefficient $\gamma$
\begin{eqnarray}
  {\bar k}_{\hbox{\scriptsize onset}}
 &=&1+{\textstyle{1\over 3-2G}}\gamma^{3/2}
     +{\textstyle{-7+2G\over (3-2G)^2}}\gamma^2 + \ldots, \nonumber\\
  \Delta_{\hbox{\scriptsize onset}} 
&=& \gamma - {\textstyle{1\over2}}\gamma^{3/2} 
	+{\textstyle{11-2G\over8(3-2G)}}\gamma^{5/2} + \ldots.
  \label{eq:threshold_ld}
\end{eqnarray}
The first correction term is proportional to $\gamma^{3/2}$, and
agrees with a low viscosity expansion of the linearized equations given by
\cite{re:muller97}. As an example, we plot in
Fig. \ref{fig:threshold} the value of the threshold,
$\Delta_{\hbox{\scriptsize onset}}$, as a function of $\gamma$ at
$G=1/3$. Previous low damping calculations of 
the standing wave amplitude equation by
\cite{re:milner91,re:miles93} and \cite{re:zhang97} considered the dominant 
term
$\Delta_{\hbox{\scriptsize onset}} = \gamma$ only. Note, however,
that the first correction, $-{1\over2}
\gamma^{3/2}$, can be a sizable contribution even for small $\gamma$
(e.g., a 15\% difference
at $\gamma=0.1$). As a reference, we note
that a similar linear analysis based on an inviscid formulation to which 
viscosity is added through a dissipation function, leads to the 
the damped Mathieu equation,
$$
  \partial_t^2 \hat{\zeta}_k(t) + \gamma\partial_t \hat{\zeta}_k(t) + 
  \omega_0^2
  (1 + 2\Delta\cos 2\omega_0t) \hat{\zeta}_k(t) = 0,
$$
where $\hat{\zeta}_{k}(t)$ is the Fourier transform of $\zeta (x,t)$.
This equation gives a threshold at $\gamma +
3\gamma^2/64+O(\gamma^3)$, which is plotted as the dot-dashed line in
Fig. \ref{fig:threshold}. The first correction term is of a
different order and has a different sign.  Finally, we mention that
rotational flow at the linear level in the surface
variables can be incorporated into the damped Mathieu equation,
as shown by \cite{re:nam93}.

\section{Standing wave amplitude Equation}
\label{sec:SWAE}

In this section, we use the multiple scale approach of \cite{re:newell69}
to derive standing wave amplitude equations valid near threshold.
It is interesting to note that the solvability condition in this
case arises from the boundary conditions, unlike
most other problems. For a driving amplitude $f$ above
threshold, we define $\epsilon = (f-f_0)/f_0$ and expand the flow as
$$
  {\bf u}=\epsilon^{1/2}{\bf u_0}+\epsilon{\bf u_1}+\epsilon^{3/2}{\bf u_2}
         +\cdots,
$$
and similarly for $p$ and $\zeta$. Near threshold, i.e., for
$\epsilon\ll 1$, we separate fast and slow time scales: $T=\epsilon
t$; $\partial_t\rightarrow\partial_t+\epsilon\partial_T$. Spatial
slow scales are not included because only regular patterns are
considered here. At order $\epsilon^{1/2}$ we
recover the linear problem discussed in the previous section. Since we are
interested in standing wave patterns of different symmetries,
the solution at this order is written as a linear combination of waves with
wavevectors ${\bf k}_m$ of magnitude $k_{\hbox{\scriptsize onset}}$ 
but along different directions on the $x$--$y$ plane,
\begin{eqnarray*}
  w_0 &=& \sum_m\cos({\bf k}_m\cdot{\bf x})B_m(T)
        \sum_{j=1,3,5,\cdots}
 e^{ji\omega t/2} w_0^j(z) e_j 
                       +\hbox{c.c.} \\
  \zeta_0 &=& \sum_m\cos({\bf k}_m\cdot{\bf x})B_m(T)
           \sum_{j=1,3,5,\cdots} e^{ji\omega t/2} e_j + \hbox{c.c.},
\end{eqnarray*}
where $B_m(T)$ are {\em real} wave amplitudes, functions only of
the slow time scale $T$, and the $e_j$ are the same as the $A_j$ found in
Eqs. (\ref{eq:relation_An}) and (\ref{eq:relation_AA}).

At order $\epsilon$ the equation of motion is
\begin{equation}
  \left(\partial_t\nabla^2 - \nu\nabla^2\nabla^2\right)w_1
     = [\nabla\times\nabla\times( {\bf u}_0 \cdot \nabla ) {\bf u}_0]
       \cdot{\bf\hat e}_z.
  \label{eq:first_order}
\end{equation}
By using the linear solution, the right hand side of
Eq. (\ref{eq:first_order}) is of the form,
\begin{equation}
  {\sum_{mn}}'\cos\big(({\bf k}_m\pm{\bf k}_n)\cdot{\bf x}\big)
	\sum_{j=0,1,2,\cdots}e^{ji\omega t}h_j(z).
  \label{eq:w_par}
\end{equation}
The first summation is over all possible ${\bf k}_m\pm{\bf k}_n$, 
except zero, and the $h_j(z)$ are combinations of exponential
functions. Since every term in the right hand side contains 
a periodic function of ${\bf x}$, and exponential functions of $t$ 
and $z$, the particular solution of Eq. (\ref{eq:first_order}),
$w_{1\hbox{\scriptsize p}}$, can be easily found. The homogeneous solution
$w_{1\hbox{\scriptsize h}}$ and $\zeta_1$ 
\begin{eqnarray}
  w_{1\hbox{\scriptsize h}} &=& {\sum_{mn}}'\cos\big(({\bf k}_m\pm{\bf k}_n)
	   \cdot{\bf x}\big)\Bigg[
         \sum_{j=1,2,3,\cdots} e^{ji\omega t}\left(
           e^{|{\bf k}_m\pm{\bf k}_n|z}\alpha_{mn}^{j\pm} 
         + e^{r_{mn}^{j\pm}z}\beta_{mn}^{j\pm}\right) + \hbox{c.c.}
   \nonumber \\
   & & \hskip200pt 
    +e^{|{\bf k}_m\pm{\bf k}_n|z}\alpha_{mn}^{0\pm}+ze^{|{\bf k}_m\pm{\bf k}_n|z}\beta_{mn}^{0\pm}\Bigg]. \nonumber\\
  \zeta_1 &=& {\sum_{mn}}'\cos\big(({\bf k}_m\pm{\bf k}_n)\cdot{\bf x}\big)
  \Bigg[
        \sum_{j=1,2,3,\cdots}e^{ji\omega t} \delta_{mn}^{j\pm} + \hbox{c.c.} 
                       + \delta_{mn}^{0\pm}\Bigg],
  \label{eq:first_order_solution}
\end{eqnarray}
must now satisfy the boundary conditions. We have defined
$\left(r_{mn}^{j\pm}\right)^2 = |{\bf k}_m\pm{\bf
k}_n|^2+ji\omega/\nu$. The constants $\alpha_{mn},\beta_{mn}$ and 
$\delta_{mn}$ are determined by the boundary conditions. At this order the
boundary conditions are,
\begin{eqnarray*}
  \partial_t\zeta_1-w_1 &=& G_{11}({\bf u_0},\zeta_0) \cr
  \left( \nabla_H^2-\partial_{zz} \right) w_1 &=& G_{12}({\bf u_0},\zeta_0) \cr
  (-\partial_t + 3\nu\nabla_H^2 + \nu\partial_{zz})\partial_zw_1
 + (g - {\sigma\over\rho}\nabla_H^2 +
    f_0\cos\omega t)\nabla_H^2\zeta_1 &=& G_{13}({\bf u_0},\zeta_0).
\end{eqnarray*}
where the functions $G_{11}$, $G_{12}$ and $G_{13}$ are listed in
appendix \ref{ap:a}. For each wavevector and harmonic (each $m$, $n$, and
$j$ in Eq. (\ref{eq:first_order_solution})), the three boundary
conditions are sufficient to determine the three unknowns 
$\alpha_{mn},\beta_{mn}$
and $\delta_{mn}$ in (\ref{eq:first_order_solution}). Because the algebra 
is quite involved (the number of terms is on the order of several thousand),
we have in practice used a symbolic manipulation package to solve for these
constants.

At order $\epsilon^{3/2}$ the equation of motion becomes,
\begin{equation}
  \left(\partial_t\nabla^2 - \nu\nabla^2\nabla^2\right)w_2
  =-\partial_T\nabla^2w_0 
   +\left\{\nabla\times\nabla\times[( {\bf u}_0 \cdot \nabla ) {\bf u}_1 
         +( {\bf u}_1 \cdot \nabla ) {\bf u}_0]\right\}\cdot{\bf\hat e}_z
  \label{eq:second_order}
\end{equation}
At this order we only need to consider resonant terms; for
example, all terms proportional to $\cos({\bf k}_1\cdot{\bf x})$. 
The right hand side of Eq.(\ref{eq:second_order}) is of the form,
\begin{equation}
  \cos({\bf k}_1\cdot{\bf x})\sum_{j=1,3,5,\cdots}e^{ji\omega t/2}E_j(z).
  \label{eq:second_order_e}
\end{equation}
where we have used the solutions $({\bf u}_0,\zeta_0,{\bf u}_1,\zeta_1)$
already determined.
Again, the $E_j(z)$ are combinations of exponential functions.
The solutions for $w_2$ and $\zeta_2$ are
\begin{eqnarray*}
  w_2 &=& \cos({\bf k}_1\cdot{\bf x}) \sum_{j=1,3,5,\cdots}
      e^{ji\omega t/2} 
      \left[\bar E_j(z) + \left(a_je^{kz}
                            + b_je^{q_jz}\right)C_j\right] \\
  \zeta_2 &=& \cos({\bf k}_1\cdot{\bf x})
      \sum_{j=1,3,5,\cdots} e^{ji\omega t/2}C_j
\end{eqnarray*}
Here $\bar E_{j}(z)$ is the particular solution that corresponds to the
right hand side at this order shown in Eq.(\ref{eq:second_order_e}),
and $a_je^{kz} + b_je^{q_jz}$ is the homogeneous solution, 
which has the same form as the linear solution. 
We now use the kinematic and tangential stress boundary conditions at this
order to determine the constants $
a_j$ and $b_j$, so that the normal stress boundary condition yields a
solvability condition for the amplitudes $C_j$, which in turn 
leads to the amplitude equations
for the $B_m$. (Note that there are various terms of $B_m$ in $\bar E_j(z)$.)

\begin{figure}
%  \centerline{\epsffile{/home/scri18/users/peilong/papers/faraday_pattern_reg/Sigma0.eps}}
\centerline{\epsffile{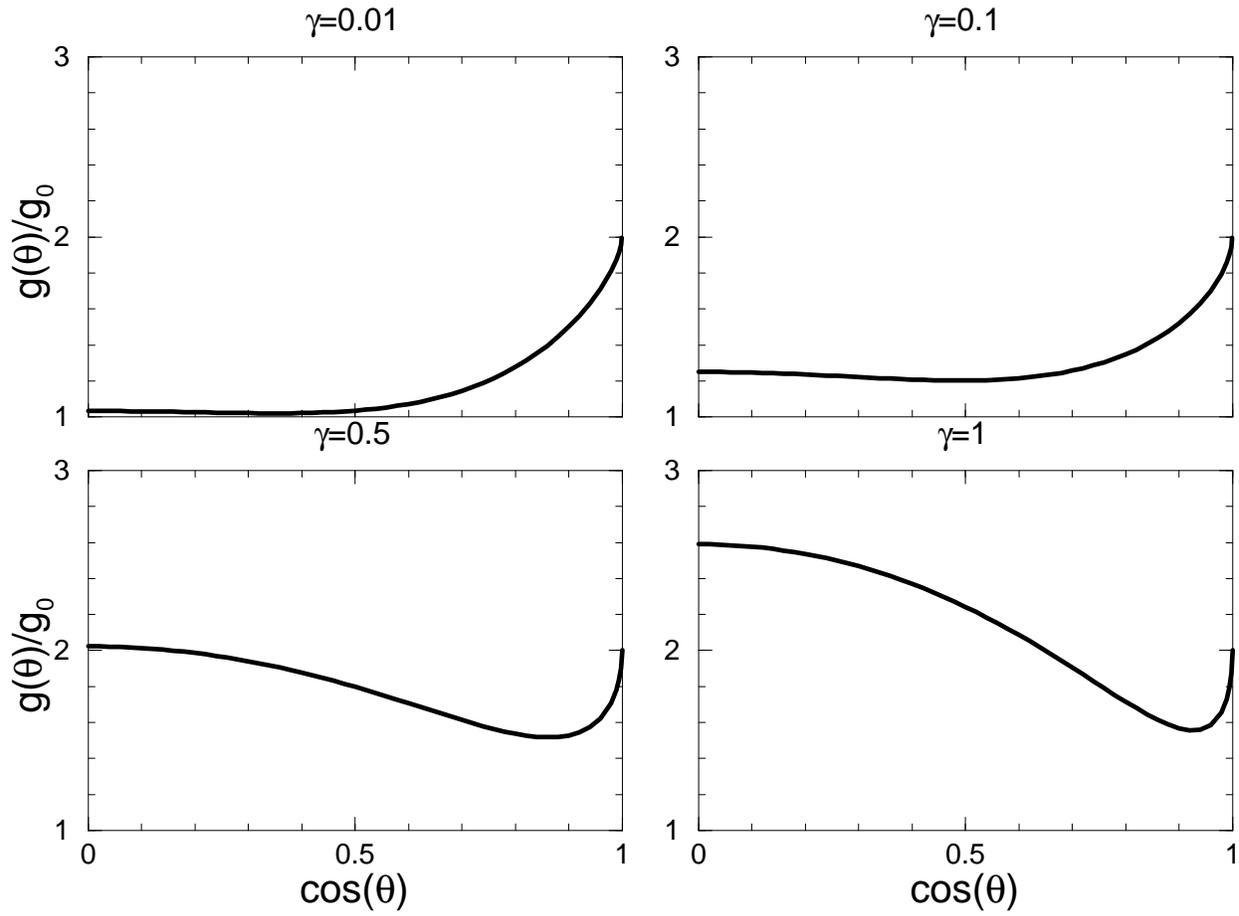}}
  \caption{Cubic term coefficient of the standing wave amplitude equation
  as a function of angle between wavevectors $\theta$, in the limit of
  gravity waves, $\Sigma=0$, and different viscous damping coefficients.}
  \label{fig:Sigma0}
\end{figure}

\begin{figure}
%  \centerline{\epsffile{/home/scri18/users/peilong/papers/faraday_pattern_reg/Sigma13.eps}}
\centerline{\epsffile{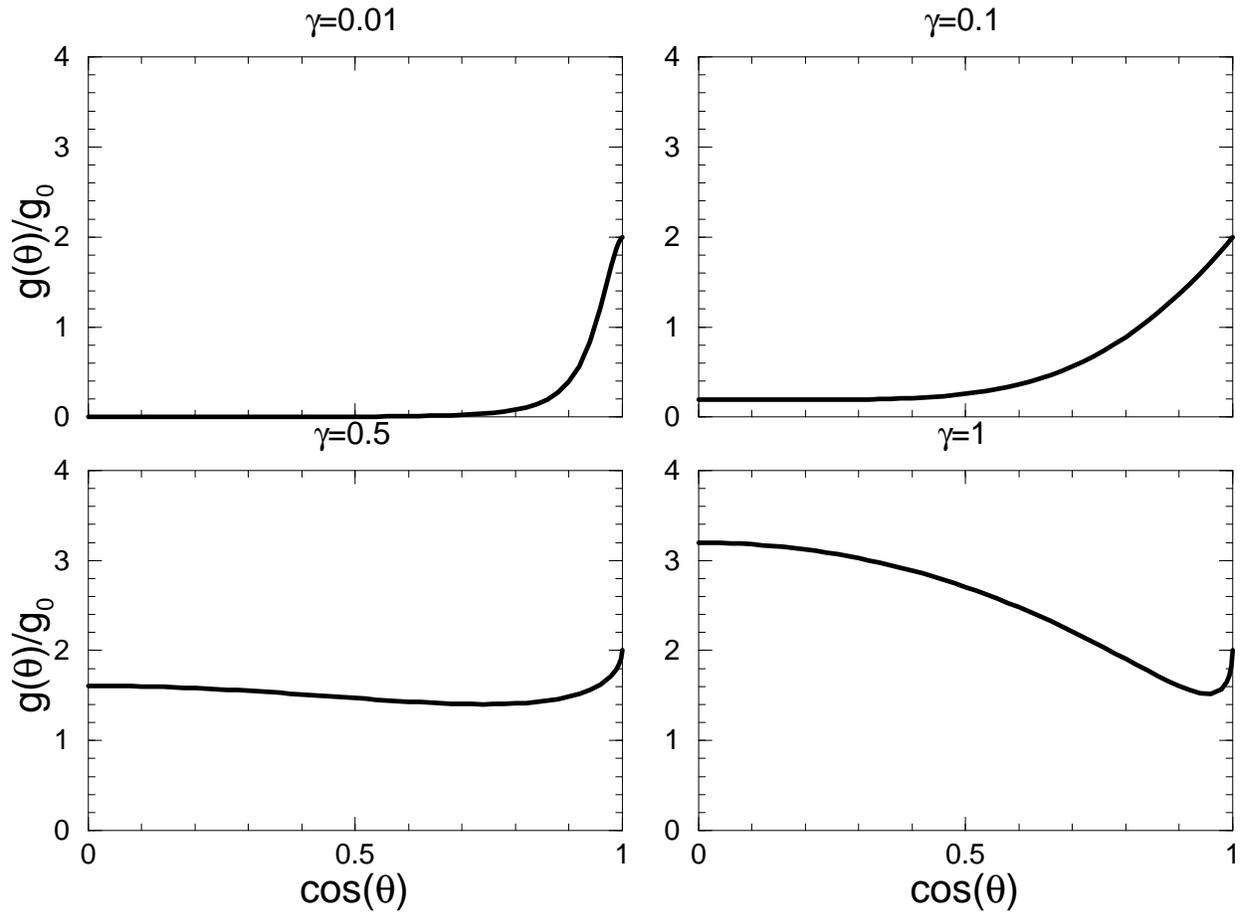}}
  \caption{Cubic term coefficient of the standing wave amplitude equation
  as a function of angle between wavevectors $\theta$, in the mixed
  capillary-gravity regime ($\Sigma=1/3$). Note that the curve becomes extremely
  flat near $\cos \theta = 0$ for low $\gamma$.}
  \label{fig:Sigma13}
\end{figure}

The boundary conditions at this order are
\begin{eqnarray*}
  \partial_t\zeta_2-w_2 &=& G_{21}({\bf u_0},\zeta_0,{\bf u_1},\zeta_1) \cr
  \left( \nabla_H^2-\partial_{zz} \right) w_2 
       	&=& G_{22}({\bf u_0},\zeta_0,{\bf u_1},\zeta_1) \cr
  (-\partial_t + 3\nu\nabla_H^2 + \nu\partial_{zz})\partial_zw_2
 + (g - {\sigma\over\rho}\nabla_H^2 + f_0\cos\omega t)\nabla_H^2\zeta_2 
	&=& G_{23}({\bf u_0},\zeta_0,{\bf u_1},\zeta_1),
\end{eqnarray*}
where the functions $G_{21}$, $G_{22}$ and $G_{23}$ are listed in 
appendix \ref{ap:a}.  By using the first two equations, $a_j$ and 
$b_j$ are found to be (again with the help of a symbolic manipulation 
package)
\begin{eqnarray*}
  a_jC_j &=& \nu(k^2+q_j^2)C_j + E^a_j \\
  b_jC_j &=& -2\nu k^2C_j + E^b_j.
\end{eqnarray*}
Here $E_j^a$ and $E_j^b$ are complicated expressions involving the amplitudes
of the waves, $B_m$. Now $w_2$ and $\zeta_2$ are substituted into the third 
boundary equation to yield
\begin{eqnarray*}
  H_1C_1 - f_0C_1^* - f_0C_3 &=& F_1, \cr
  H_3C_3 - f_0C_1 - f_0C_5 &=& F_3, \cr
  H_5C_5 - f_0C_3 - f_0C_7 &=& F_5, ~~ \ldots \cr
\end{eqnarray*}
The left hand side of this system of equations is identical to 
Eq. (\ref{eq:eigen_zero}) for the linear
problem, and the functions $F_j$ on the right hand side are functions 
of $B_m$ and $dB_1/dT$.  Solving for $C_j$ just like we solved for the 
linear threshold, we obtain
$$
  \bar H_1C_1 - f_0 C_1^* = F_1 + {f_0\over \bar H_3}\left(F_3
      +{f_0\over \bar H_5}(F_5 + \cdots)\right) \equiv F,
$$
with $\bar H_j$ defined similarly to $\bar H_1$
in Eq. (\ref{eq:relation_AA}). Since the threshold of linear instability 
given by $f_0=|\bar H_1|$, a nontrivial solution for $C_1$ will exist if the
following solvability condition is satisfied:
$$
  F\bar H_1^*+F^*f_0=0.
$$ 
This condition immediately leads to a standing wave amplitude equation 
for $B_1$,
\begin{equation}
  {dB_1\over dT} = \alpha B_1 - g_0 B_1^3
   - \sum_{m\not=1}g(\theta_{m1})B_m^2B_1,
  \label{eq:amplitude_equation}
\end{equation}
with $\theta_{m1}$ the angle between ${\bf k}_m$ and ${\bf k}_1$.
The linear coefficient $\alpha$ (times $\epsilon$) is the linear growth or 
decay rate of this wave, and can be obtained from the linear analysis
(simply consider an extra factor $e^{\alpha t}$ in
Eq. (\ref{eq:linear_solution})).
The coefficient $g(\theta)$ describes the nonlinear interaction
between different linearly unstable modes, and provides for
the saturation of the wave amplitude. Figures \ref{fig:Sigma0} and 
\ref{fig:Sigma13} show our results for different values of $\gamma$
and for $\Sigma = 0$ (pure gravity waves), and $\Sigma = 1/3$ (mixed
gravity-capillary waves). It is also important to note the 
asymptotic behavior of $g(\theta)$ as $\nu \rightarrow 0$.
We have already discussed in Sec. \ref{sec:equations},
that the irrotational component of the flow contributes to order $\nu$
to the linearized equation of motion (Eq. (\ref{eq:w0linear})), whereas
the rotational flow contribution scales as $\nu^{3/2}$ instead. This
observation is the basis for earlier low viscosity approximations 
in which only viscous dissipation arising from the irrotational flow was
considered (\cite{re:miles84,re:milner91,re:miles93}).
However, we have computed the coefficient $g(\theta)$ with and without
the linear rotational flow and observed that both contributions are of
order $\nu$ at small $\nu$. Therefore, a formulation that does not
incorporate the rotational flow explicitly cannot obtain the correct form
of the third order damping coefficients, even in the limit of small
viscosity.

\begin{figure}
%  \epsffile{/home/scri18/users/peilong/papers/faraday_pattern_reg/fig3.ps}
\epsffile{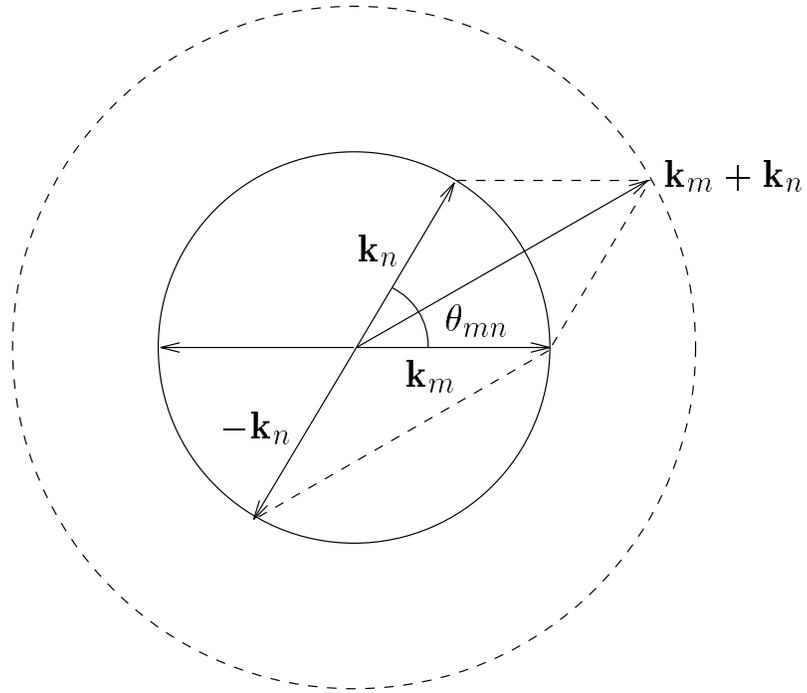}
  \caption{Schematic representation of a triad resonant interaction: two
  linearly unstable modes ${\bf k}_{m}$ and ${\bf k}_{n}$ interact to produce a
  linearly stable mode. This mode interacts with $-{\bf k}_{n}$ leading to
  resonance with ${\bf k}_{m}$.}
  \label{fig:triad}
\end{figure}

Particular nonlinear interaction terms that contribute to $g(\theta)$
are shown in Fig. \ref{fig:triad}. Two linearly
unstable modes with wave vectors ${\bf k}_m$ and ${\bf
k}_n$ ($|{\bf k}_m|$ $=$ $|{\bf k}_n|$ $=$ $k_{\hbox{\scriptsize
onset}}$) interact to produce a wave at ${\bf k}_m+{\bf k}_n$
with an amplitude proportional to $B_mB_n$. This mode corresponds
to a first order solution ($w_1$ and $\zeta_1$ in
Eqs. (\ref{eq:w_par}) and Eq. (\ref{eq:first_order_solution})). Now
${\bf k}_m+{\bf k}_n$ couples back to the original wave at 
$-{\bf k}_n$ to give a contribution $B_n^2B_m$ to
$dB_m/dT$. Since the mode ${\bf k}_m+{\bf k}_n$ is damped (only
waves with wavenumbers near $k_{\hbox{\scriptsize onset}}$ are
unstable), this is a dissipative term and contributes to
nonlinear saturation of the wave. Triad resonance occurs when
the frequency of the mode ${\bf k}_m+{\bf k}_n$ equals the driving
frequency (the modes ${\bf k}_m$ and ${\bf k}_n$ oscillate at half the
driving frequency). Energy is now directly transferred into this
mode which can have a very large amplitude at low damping.
Since ${\bf k}_m+{\bf k}_n$ couples back to $-{\bf k}_n$, it provides 
a dissipation channel for the mode ${\bf k}_n$. Dissipation is
enhanced by triad resonance and results in a large value of 
$g(\theta_{mn})$ in the vicinity of the resonant angle. The
resonant angle can be estimated from the inviscid dispersion relation
(\ref{eq:ideal_dispersion}), written in dimensionless form,
\begin{equation}
  \bar\omega^2 = G\bar k + \Sigma\bar k^3,
  \label{eq:ideal_dispersion_d}
\end{equation}
with $\bar k=1$, and $\bar\omega^2 = G+\Sigma = 1$ for the linearly
unstable mode. At resonance, we have $\bar\omega = 2$, and the
resonant wave number $\bar k_r=|\bar{\bf k}_m+\bar{\bf k}_n|$
satisfies $\bar k_r\left( G+\Sigma\bar k_r^2\right) = 4$.
If $\theta_r$ is the resonant angle between ${\bf k}_m$ and ${\bf
k}_n$, $\bar k_r = \sqrt{2(1+\cos\theta_r)}$, the resonance condition
becomes
\begin{equation}
  \sqrt{2(1+\cos\theta_r)}[G+2(1+\cos\theta_r)\Sigma] = 4.
  \label{eq:resonance_angle}
\end{equation}
Because $G+\Sigma=1$, this condition can only be satisfied when
$\Sigma > 1/3$. For example, $\cos\theta_r = 2^{1/3}-1$ for $\Sigma = 1$. 

\begin{figure}
%  \centerline{\epsffile{/home/scri18/users/peilong/papers/faraday_pattern_reg/Sigma1.eps}}
\centerline{\epsffile{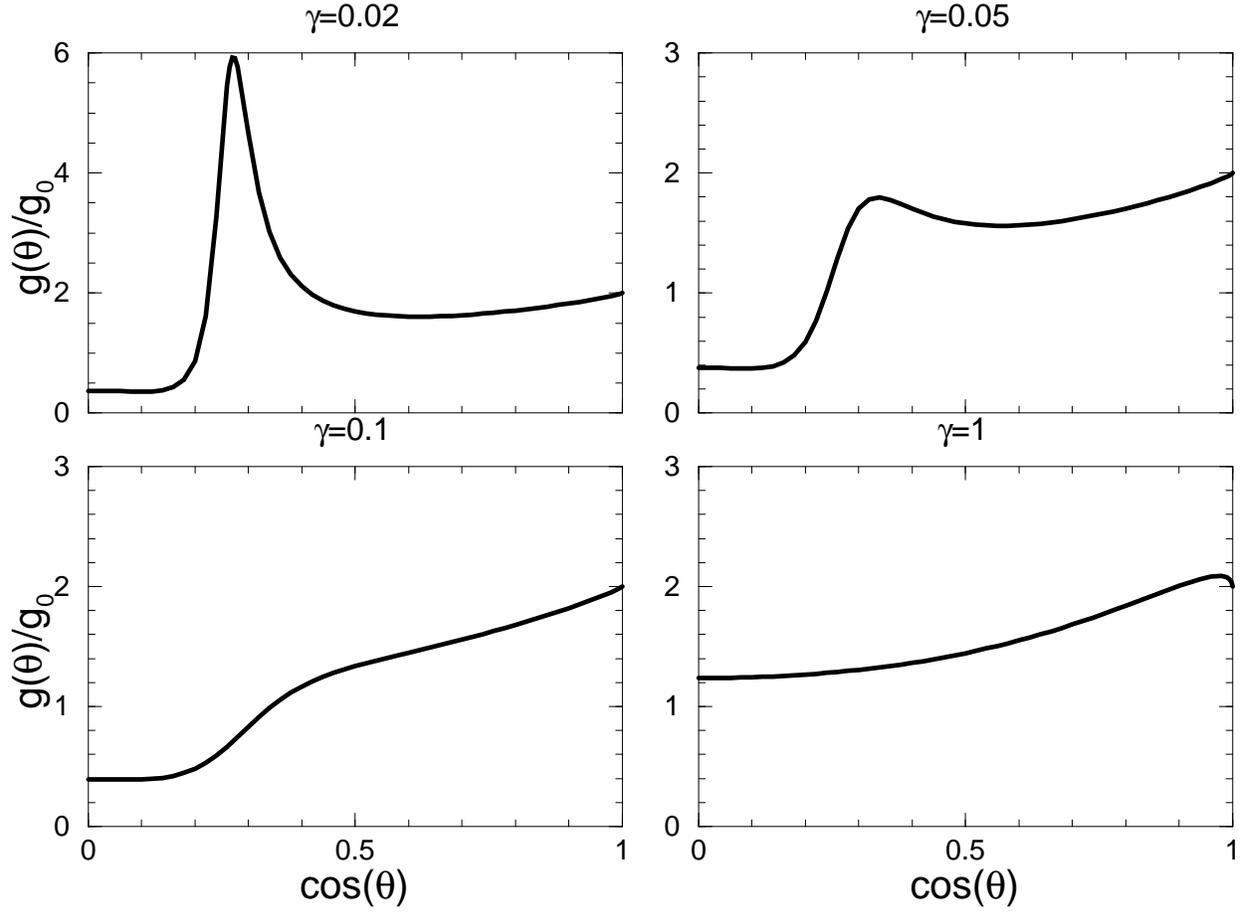}}
  \caption{Cubic term coefficient of the standing wave amplitude equation
  as a function of angle between wavevectors $\theta$, in the limit of
  capillary waves, $\Sigma=1$. The large peaks at small values of $\gamma$ are
  due to triad resonant interactions.}
  \label{fig:Sigma1}
\end{figure}

For finite damping, the resonance condition is modified. However,
triad resonance is expected to be significant only at low damping because of 
the damped nature of the first order wave. For example, Fig. \ref{fig:Sigma1}
shows $g(\theta)$ for different $\gamma$ and $\Sigma=1$. At small
$\gamma$, the nonlinear coefficient grows near resonance and peaks at the 
resonant angle. The value of the peak is seen to decrease with increasing 
$\gamma$. At $\gamma = 0.1$, resonance has almost disappeared.

In the formulation presented earlier, resonance arises from the homogeneous
solutions $w_{1\hbox{\scriptsize h}}$ and $\zeta_1$, which require
finding the constants $\alpha_{mn},\beta_{mn}$ and $\delta_{mn}$ in
Eq. (\ref{eq:first_order_solution}) by enforcing the boundary conditions
at first order. The
boundary conditions give rise to a system of linear equations for
$\alpha_{mn},\beta_{mn}$ and $\delta_{mn}$, the left hand side 
of which (its matrix form is explicitly given in appendix
\ref{ap:b}) at $\gamma=0$ has a determinant
$$
  8\bar k^2\left(G\bar k^2 + \Sigma\bar k^4\right)
  \left[4\bar k - \left(G\bar k^2+\Sigma\bar k^4\right)\right]^2,
$$
which, when equated to zero, is equivalent to Eq. (\ref{eq:resonance_angle}).

\section{Pattern selection and comparison with experiments}

Since the standing wave amplitude equation (\ref{eq:amplitude_equation}) can 
be written in gradient form, the selected pattern near threshold
immediately follows
(\cite{re:cross93}). Equation (\ref{eq:amplitude_equation}) is equivalent to
$$
  {dB_n\over dT} = - {\delta{\cal F}\over \delta B_n},
$$
with the Lyapunov function ${\cal F}$ given by
$$
  {\cal F} =-{1\over2}\alpha\sum_mB_m^2
            +{1\over4}\sum_m\sum_n g(\theta_{mn})B_m^2B_n^2,
$$
with $g_{0} = g(\theta_{nn}) $ which equals half the value of
$g(\theta\rightarrow 0)$. The amplitude equation then implies that
$$
  {d{\cal F}\over dT} = \sum_n{\delta{\cal F}\over\delta B_n}
                              {dB_n\over dT}
                      = - \sum_n \left({dB_n\over dT}\right)^2 \le 0,
$$
so that the preferred pattern can be determined by minimization of
${\cal F}$.  The experimentally observed regular patterns
above onset consist of $N$ standing waves, with uniform amplitudes and
wavevectors ${\bf k}_m, m=1 \ldots N$. The case $N=1$
corresponds to a single standing waves (a pattern of parallel stripes), 
$N=2$ to a pattern of square symmetry, $N=3$ of hexagonal symmetry,
etc. For these regular patterns, the standing wave amplitudes are
$$
  B_n^2 = {\alpha\over g_0 + \sum_{m\not=n}g(\theta_{mn})},
  \qquad n=1\cdots N.
$$
The value of the Lyapunov function as a function of $N$ then becomes
\begin{equation}
  {\cal F}(N) = -{\alpha^2\over4}{N\over g_0+\sum_{m=2}^Ng(\theta_{m1})}.
  \label{eq:lyapunov}
\end{equation}

\begin{figure}
%  \centerline{\epsffile{/home/scri18/users/peilong/papers/faraday_pattern_reg/lya.eps}}
\centerline{\epsffile{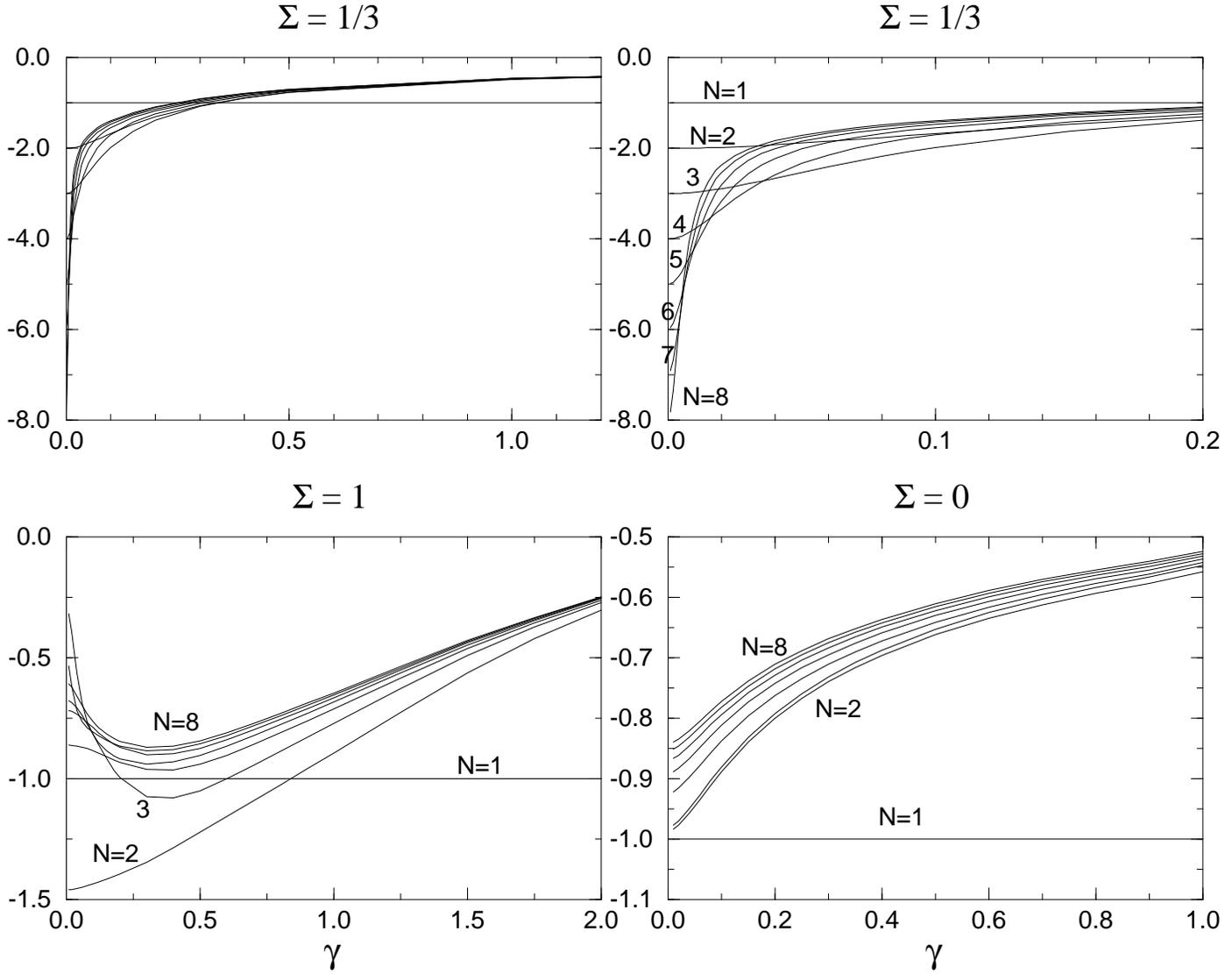}}
  \caption{Numerical values of the Lyapunov function for regular patterns
  comprising $N$ standing waves as a function of the viscous damping parameter
  $\gamma$. Bottom right: gravity wave limit; bottom left: capillary wave limit;
  top left, the mixed case of $\Sigma = 1/3$; top right is the same as top left
  but showing the region of small damping in more detail. In the two bottom
  plots, the curves not labeled are ordered in increasing order of $N$.}
  \label{fig:lya}
\end{figure}

Figure \ref{fig:lya} shows the computed values of ${\cal F}(N)$ as a 
function of $\gamma$ for different values of $\Sigma$. For pure gravity waves
($\Sigma=0$), the $N=1$ state has the lowest value of the
Lyapunov function and hence will be the selected pattern. 
At low frequency, the system effectively crosses over to the large
damping region regardless of its (finite) viscosity (this range was not
accessible to the low damping calculation of
of \cite{re:zhang97}). On the other hand, for
pure capillary waves ($\Sigma=1$) the preferred pattern is $N=2$ at
low damping and $N=1$ at high damping. 
Interesting behavior is observed in the vicinity of $\Sigma=1/3$ 
(mixed gravity capillary waves) where the triad resonance angle
approaches zero. Hexagonal and higher symmetry quasipatterns are selected
with decreasing $\gamma$. The low damping results in this region 
are in qualitative agreement with the earlier work of 
\cite{re:zhang97}, although
the latter could not account for the transition to $N=1$ as $\gamma$ is
increased.

We finally compare our predictions (based on Eq. (\ref{eq:lyapunov})) and two
recent sets of experiments that addressed pattern selection in
the large aspect
ratio limit by \cite{re:kudrolli96a} and by \cite{re:binks97}. The only 
input parameters in our calculations are the fluid properties (density, 
surface 
tension and viscosity), and the frequency of the driving acceleration. All 
these parameters are known fairly precisely in the experimental work, thus 
allowing a quantitative comparison between theory and experiments.

Pattern selection in the low viscosity range has been recently studied by
\cite{re:binks97}. They have developed a cell of exceptionally large aspect 
ratio, and of depth that is much larger than the wavelength. 
The fluid used was a low viscosity, low surface tension silicon oil with
$\nu = 0.03397$~cm$^{2}$/s, $\rho$ = 0.8924~g/cm$^{3}$ and $\sigma$ = 
18.3~dyne/cm. Given the range of frequencies studied, the
viscous damping parameter probed was within $\gamma \sim 0.01 - 0.03$.
They have reported transitions from standing wave patterns of square symmetry 
at high frequency (> 41~Hz), to hexagonal, eight-fold and ten-fold 
quasi-periodic patterns upon lowering the driving frequency.
Stable hexagonal patterns appear at approximately 36~Hz, although a 
transition region of mixed square/hexagonal symmetry is observed between 
approximately 36~Hz and 41~Hz. Given the parameters of this
experiment, our theory predicts a transition at 35.4~Hz, compared to
the value of 32.8~Hz given by the earlier work of \cite{re:zhang97}.
An additional transition region exhibiting patterns of mixed hexagonal
and eight-fold symmetry was also observed between 30-31~Hz, which
compares favorably with our prediction for the transition to eight-fold
symmetric patterns at 28.7~Hz (\cite{re:zhang97} had predicted the
transition to occur at 27.9~Hz).
\footnote{
In this experiment, patterns with $N=5$ are
observed around 27Hz. We agree with the authors of the experiment that
this discrepancy may be due to finite size effects. In fact,
${\cal F}(5)$ is very close to ${\cal F}(4)$ at about
26Hz (the difference is 
less than $0.2\%$), although ${\cal F}(5)>{\cal F}(4)$.
}
A possible explanation for the larger discrepancy between the experiments and
the calculations of \cite{re:zhang97} involves the fact that they 
only used the term linear in $\gamma$ in the calculation of the 
threshold for instability (Eq. (\ref{eq:threshold_ld})). Omitting
the first correction (of order $\gamma^{3/2}$) 
yields a similar percentage error in the threshold value
(for $\gamma$ in the range 0.01 -- 0.03  as is 
appropriate for this experiment).

\begin{figure}
%  \centerline{\epsffile{/home/scri18/users/peilong/papers/faraday_pattern_reg/expt.eps}}
\centerline{\epsffile{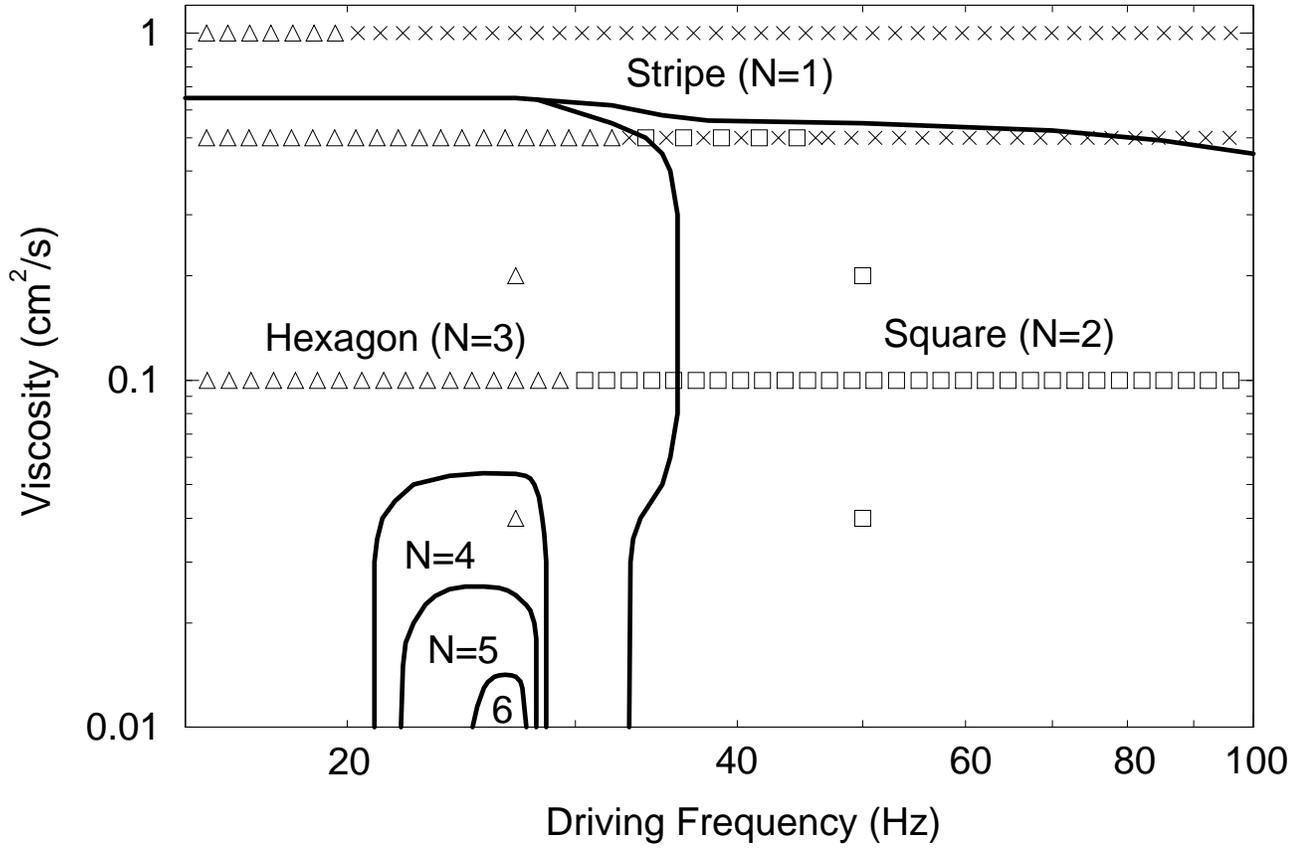}}
  \caption{Selected patterns as a function of fluid viscosity and driving 
  frequency. The symbols are the experimental results of 
  \cite{re:kudrolli96a}. $\times$, stripe patterns; $\Box$, square
  patterns; and, $\triangle$, hexagonal patterns. Alternating $\times$ and 
  $\Box$ indicate regions in which stationary mixed stripe and square patterns
  were observed.}
  \label{fig:expt}
\end{figure}

Another set of recent experiments involving fluids of different
viscosity has been carried out by
\cite{re:kudrolli96a}. Although the depth of the fluid layer (0.3~cm) is 
smaller than the wavelength of the waves (1--3~cm), the comparison is 
still illuminating. Figure \ref{fig:expt} shows the symmetry of the predicted 
patterns as a function of the viscosity of the fluid and of the driving 
frequency (with $\rho=0.95\hbox{g}/\hbox{cm}^3$ and
$\sigma=20.6\hbox{dyn}/\hbox{cm}$), and the experimentally observed
patterns. They find a stripe pattern at high viscosity,
a hexagonal pattern at low viscosity and frequency, and a square
pattern at low
viscosity and high frequency. Two significant discrepancies concern the
experimental observation of a hexagonal pattern at
$\nu=1\hbox{cm}^2/\hbox{s}$ and low frequency, and also at 
$\nu=0.04\hbox{cm}^2/\hbox{s}$ and $f=27\hbox{Hz}$. It is possible that the
shallowness of the fluid layer can account for these differences, especially 
in view of the fact that, as noted above, the experiments by \cite{re:binks97} 
did probe this latter region in a deep fluid layer, and their results 
do agree with our predictions.

Finally, the fact that portions of the
boundaries separating regions of different symmetry appear almost as straight 
lines in Fig. \ref{fig:expt} is due to the log-log scale used. In addition,
the transition line delimiting the region of stripe patterns is almost
independent of frequency only because of the limited frequency range probed in
the experiments and displayed in the figure. On the other hand, 
the line separating regions of square and
hexagonal patterns is almost independent of viscosity because it depends mainly
on whether the waves are capillarity or gravity dominated, fact that is largely
dependent on the driving frequency and not on viscosity.

In summary, we have presented a nonlinear theory for Faraday waves
in viscous fluids with no assumptions or approximations other than
those inherent to the multi-scale expansion. A set of standing wave
amplitude equations has been obtained that is of gradient
form. Minimization of the associated Lyapunov function leads to
determination of the preferred pattern near threshold. The predicted
patterns are in excellent agreement with recent experiments in large
aspect ratio systems involving a range of fluid viscosities and
driving frequencies. According to Fig. \ref{fig:lya}, the transition from 
square to stripe patterns remains in the capillary wave limit 
of $\Sigma = 1$ (high frequency
limit in the experiments). However, the figure for $\Sigma=0$ indicates that
stripe patterns are always preferred in the pure-gravity-wave
limit (low frequency limit in the experiments). 
Furthermore, all the high symmetry patterns 
(with $N\geq 3$) are observed in the vicinity of $\Sigma=1/3$,
point at which the triad resonant angle approaches zero, and for low damping
where the resonance is more pronounced. 

\begin{acknowledgments}
This research has been supported by the U.S.  Department of Energy, contract 
No.  DE-FG05-95ER14566, and also in part by the Supercomputer Computations 
Research Institute, which is partially funded by the U.S.  Department of 
Energy, contract No. DE-FC05-85ER25000.
\end{acknowledgments}

\newpage
\appendix

\section{Inhomogeneous terms of the first and second order equations}
\label{ap:a}

We list in this appendix the functions $G_{ij}$, the inhomogeneous terms in the
boundary conditions at first and second order.
\begin{eqnarray*}
  G_{11} &=& \partial_z w_0\zeta_0 
   - u_0\partial_x\zeta_0 - v_0\partial_y\zeta_0, \cr
  G_{12} 
    &=& \partial_x\left[ -\partial_{zz}u_0\zeta_0-\partial_{xz}w_0\zeta_0
                  +2(\partial_x u_0-\partial_z w_0) \partial_x\zeta_0
                  +(\partial_y u_0+\partial_x v_0)\partial_y\zeta_0 \right] \\
    &+& \partial_y\left[ -\partial_{zz}v_0\zeta_0-\partial_{yz}w_0\zeta_0
                  +2(\partial_y v_0-\partial_z w_0)\partial_y\zeta_0
                  +(\partial_x v_0+\partial_y u_0)\partial_x\zeta_0 \right],\cr
  G_{13}
    &=& -\rho\nabla_H\cdot[({\bf u}_0\cdot\nabla){\bf u}_0]
       +\nabla_H^2\left(-2\eta\partial_{zz}w_0\zeta_0
                          + \partial_zp_0\zeta_0 \right) \cr
  G_{21} &=& - \partial_T\zeta_0 
        - u_0\partial_x\zeta_1 - u_1\partial_x\zeta_0 
        - \partial_z u_0\zeta_0\partial_x\zeta_0 
        - v_0\partial_y\zeta_1 - v_1\partial_y\zeta_0 
        - \partial_z v_0\zeta_0\partial_y\zeta_0 \cr 
     &+& \partial_z w_0\zeta_1 + \partial_z w_1\zeta_0
        +{\textstyle{1\over2}}\partial_{zz}w_0\zeta_0^2, \cr
  G_{22} &=& \partial_x\Big[
    - \partial_{zz}u_1\zeta_0 - \partial_{zz}u_0\zeta_1
    - {\textstyle{1\over2}}\partial_{zzz}u_0\zeta_0^2
    - \partial_{xz}w_1\zeta_0 - \partial_{xz}w_0\zeta_1
    - {\textstyle{1\over2}}\partial_{xzz}w_0\zeta_0^2 \\
  &-& 2(\partial_z w_1-\partial_x u_1)\partial_x\zeta_0 
    - 2(\partial_z w_0-\partial_x u_0)\partial_x\zeta_1
    - 2\partial_z(\partial_z w_0-\partial_x u_0)\zeta_0\partial_x\zeta_0 \\
  &+& (\partial_y u_1+\partial_x v_1)\partial_y\zeta_0
    + (\partial_y u_0+\partial_x v_0)\partial_y\zeta_1
    + \partial_z(\partial_y u_0+\partial_x v_0)\zeta_0\partial_y\zeta_0\Big] \\
  &+& \partial_y\Big[
    - \partial_{zz}v_1\zeta_0 - \partial_{zz}v_0\zeta_1
    - {\textstyle{1\over2}}\partial_{zzz}v_0\zeta_0^2
    - \partial_{yz}w_1\zeta_0 - \partial_{yz}w_0\zeta_1
    - {\textstyle{1\over2}}\partial_{yzz}w_0\zeta_0^2 \\
  &-& 2(\partial_z w_1-\partial_y v_1)\partial_y\zeta_0 
    - 2(\partial_z w_0-\partial_y v_0)\partial_y\zeta_1
    - 2\partial_z(\partial_z w_0-\partial_y v_0)\zeta_0\partial_y\zeta_0 \\
  &+& (\partial_y u_1+\partial_x v_1)\partial_x\zeta_0
    + (\partial_y u_0+\partial_x v_0)\partial_x\zeta_1
    + \partial_z(\partial_y u_0+\partial_x v_0)\zeta_0\partial_x\zeta_0\Big] \\
  G_{23} &=& 
    - \rho\nabla_H\cdot[({\bf u}_0\cdot\nabla){\bf u}_1 +({\bf
	  u}_1\cdot\nabla{\bf u}_0)]
    + \rho\partial_T\partial_z w_0
    + \nabla_H^2\big[ -{\textstyle{1\over2}}\rho f_0\left(e^{i\omega t}
				+e^{-i\omega t}\right)\zeta_0 \\
  &+& \partial_z p_1\zeta_0 + \partial_z p_0\zeta_1
    + {\textstyle{1\over2}}\partial_z^2p_0\zeta_0^2
    - 2\eta\partial_z^2w_1\zeta_0 - 2\eta\partial_z^2w_0\zeta_1
    - \eta\partial_z^3w_0\zeta_0^2 \\
  &+& 2\eta(\partial_zu_1+\partial_xw_1)\partial_x\zeta_0
    + 2\eta(\partial_zw_0-\partial_xu_0)(\partial_x\zeta_0)^2 \\
  &+& 2\eta(\partial_zv_1+\partial_yw_1)\partial_y\zeta_0
    + 2\eta(\partial_zw_0-\partial_yv_0)(\partial_y\zeta_0)^2 \\
  &-& 2\eta(\partial_yu_0+\partial_xv_0)\partial_x\zeta_0\partial_y\zeta_0
    - {\textstyle{3\over2}}\sigma\partial_{xx}\zeta_0(\partial_x\zeta_0)^2
    - {\textstyle{3\over2}}\sigma\partial_{yy}\zeta_0(\partial_y\zeta_0)^2 \\
 &-& {\textstyle{1\over2}}\sigma\partial_{xx}\zeta_0(\partial_y\zeta_0)^2
   -{\textstyle{1\over2}}\sigma\partial_{yy}\zeta_0(\partial_x\zeta_0)^2
   -2\sigma\partial_x\zeta_0\partial_y\zeta_0
         \partial_{xy}\zeta_0 \big].
\end{eqnarray*}

\newpage
\section{Matrix of coefficients at first order}
\label{ap:b}

Left hand side of the system of linear equations for the first
order solution (for simplicity, we only show the case
$\gamma\ll 1$ and the coefficients of first time harmonic $e^{i\omega t/2}$),
$$
  \left(
  \begin{array}{ccccccccc}
     -1 & 0 & 0 & -1 & 0 & 0 & -2i & 0 & 0 \\
      0 & -1 & 0 & 0 & -1 & 0 & 0 & 0 & 0 \\
      0 & 0 & -1 & 0 & 0 & -1 & 0 & 0 & 2i \\
     -\gamma\bar k^2 & 0 & 0 & 2i & 0 & 0 & 0 & 0 & 0 \\
      0 & -\bar k^2 & 0 & 0 & -\bar k^2 & 0 & 0 & 0 & 0 \\
      0 & 0 & -\gamma\bar k^2 & 0 & 0 & -2i & 0 & 0 & 0 \\
     -2i\bar k & 0 & 0 & \gamma\bar k^2\bar q^* & 0 & 0 & G\bar k^2 + \Sigma\bar k^4 & 2\gamma\bar k^2 & 0 \\
      0 & \gamma\bar k^3 & 0 & 0 & 0 & 0 & 2\gamma\bar k^2 & G\bar k^2 + \Sigma\bar k^4 & 2\gamma\bar k^2 \\
      0 & 0 & 2i\bar k & 0 & 0 & \gamma\bar k^2\bar q & 0 & 2\gamma\bar k^2 & G\bar k^2 + \Sigma\bar k^4
  \end{array} \right) \left(
  \begin{array}{l}
    \alpha_{mn}^{1-} \\ \alpha_{mn}^0 \\ \alpha_{mn}^{1+} \\
    \beta_{mn}^{1-} \\ \beta_{mn}^0 \\ \beta_{mn}^{1+} \\
    \delta_{mn}^{1-} \\ \delta_{mn}^0 \\ \delta_{mn}^{1+}
  \end{array}
  \right)
$$
with $\bar k = |\bar{\bf k}_m+\bar{\bf k}_n|$ and $\bar q^2 \equiv
\bar k^2 + 2i/\gamma$.

\newpage
% \begin{table}
% \caption{Frequencies delimiting regions in which patterns of different 
% symmetries are selected for
% $\nu=0.03397\hbox{cm}^2/\hbox{s}$ ($\gamma\sim 0.01 - 0.03$),
% $\rho=0.8924\hbox{g}/\hbox{cm}^3$, and $\sigma=18.3\hbox{dyne}/\hbox{cm}$.}
% \label{table1}
% \vspace{0.5cm}
% \begin{tabular}{cccc}
% Pattern & Experiment    & Theory & Weak-damping theory \\
% transition & \cite{re:binks97}  &       &\cite{re:zhang97}\\
%            &   (Hz)    &  (Hz)       &  (Hz)   \\
% \\
% $2\leftrightarrow3$ & 35 & 35.4 & 32.8 \\
% $3\leftrightarrow4$ & 29 & 28.7 & 27.9 \\
% \end{tabular}
% \end{table}

% \newpage

\vskip 5cm

\end{document}